\documentclass[preprint2]{aastex61}

\usepackage{url}

\shorttitle{Automatized photometric monitoring of AGNs}
\shortauthors{F. Pozo Nu\~nez et al.}

\begin{document}

\title{Automatized photometric monitoring of active galactic nuclei with the 46 cm telescope of the Wise observatory}

\correspondingauthor{Francisco Pozo Nu\~nez}
\email{francisco.pozon@gmail.com}

\author[0000-0002-6716-4179]{Francisco Pozo Nu\~nez}
\affiliation{Department of Physics, Faculty of Natural Sciences, University of Haifa, Haifa 31905, Israel.}
\affiliation{Haifa Research Center for Theoretical Physics and Astrophysics, Haifa 31905, Israel.}

\author{Doron Chelouche}
\affiliation{Department of Physics, Faculty of Natural Sciences, University of Haifa, Haifa 31905, Israel.}
\affiliation{Haifa Research Center for Theoretical Physics and Astrophysics, Haifa 31905, Israel.}

\author{Shai Kaspi}
\affiliation{Wise Observatory School of Physics and Astronomy, Tel Aviv University, Tel Aviv 69978, Israel.}

\author{Saar Niv}
\affiliation{Department of Physics, Faculty of Natural Sciences, University of Haifa, Haifa 31905, Israel.}

\begin{abstract}

We present the first results of an ongoing variability monitoring program of active galactic nuclei (AGNs) using the 46\,cm telescope of the Wise observatory in Israel. The telescope has a field of view of $1.25^{\circ} \times 0.84^{\circ}$ and is specially equipped with five narrow band filters at 4300, 5200, 5700, 6200 and 7000\,\AA\, in order to perform photometric reverberation mapping studies of the central engine of AGNs. The program aims to observe a sample of 27 AGNs (V $<17$ mag) selected according to tentative continuum and line time delay measurements obtained in previous works. We describe the autonomous operation of the telescope together with the fully automatic pipeline used to achieve high-performance unassisted observations, data reduction, and light curves extraction using different photometric methods. The science verification data presented here, demonstrates the performance of the monitoring program in particular for efficiently photometric reverberation mapping of AGNs with additional capabilities to carry out complementary studies of other transient and variable phenomena such as variable stars studies.

\end{abstract}




\section{Introduction} \label{sec:intro}

Variability studies have been providing unique information about the inner structure of the central region of active galactic nuclei (AGNs) over the last five decades. It is now generally accepted that AGNs are composed of a central super-massive black hole with an accretion disk (AD). The AD generates the blue/UV continuum radiation, which is highly variable on different time scales (see reviews by \citealt{2015arXiv150102001A} and \citealt{2015ARA&A..53..365N}).

As the central engine of AGNs cannot be resolved by conventional imaging techniques, the only method available, which is independent of spatial resolution, is reverberation mapping (\citealt{1973ApL....13..165C}; \citealt{1982ApJ...255..419B}; \citealt{1986ApJ...305..175G}). Reverberation mapping uses spectroscopic (e.g. \citealt{1993PASP..105..247P}; \citealt{2004ApJ...613..682P}; \citealt{2012ApJ...755...60G}) and photometric (\citealt{2011A&A...535A..73H}; \citealt{{2012A&A...545A..84P}}; \citealt{2012ApJ...747...62C}) monitoring to measure the time delay, $\tau$, between changes in the optical-ultraviolet continuum produced in the hot AD and the emission lines from the broad line region (BLR) gas clouds or the molecular dust torus, thereby estimating the effective BLR radius, $R_{BLR}=c \cdot \tau_{BLR}$ ($c$ is the speed of light) and the dust torus size $R_{dust}=c \cdot \tau_{dust}$ (e.g. \citealt{2006ApJ...639...46S}; \citealt{2014ApJ...788..159K}; \citealt{2014A&A...561L...8P}). 

A further use of spectroscopic and photometric reverberation mapping is to map the AD. Depending on the local temperature of the AD, the time delays between light curves at different continuum bands can be interpreted as the light travel time ($\tau_{AD}$) across the AD (e.g. \citealt{1998ApJ...500..162C}). Therefore, as a first approximation, the time delays yield valuable information about the size ($R_{AD} \sim c \cdot \tau_{AD}$) and the temperature stratification across the AD, both crucial parameters provide important constraints on actual theoretical models of the AD (e.g. \citealt{2013ApJ...772....9C}, and references therein).

While great observational progress has been made, time delays between different continuum bands have been detected for only a few AGNs over the past years (e.g. \citealt{2001MNRAS.325.1527C}; \citealt{2005ApJ...622..129S}; \citealt{2007MNRAS.380..669C}; \citealt{2009A&A...493..907B}; \citealt{2011MNRAS.415.1290L}; \citealt{2014ApJ...788...48S}; \citealt{2016MNRAS.456.4040T}; \citealt{2017ApJ...840...41E}). For some objects the measured time delay shows large uncertainties, likely due to under-sampled light curves and contribution of the BLR emission lines to the bands, hence photometric monitoring with higher precision are highly desired.

\begin{figure}[t!]
  \centering
  \includegraphics[angle=0,width=\columnwidth]{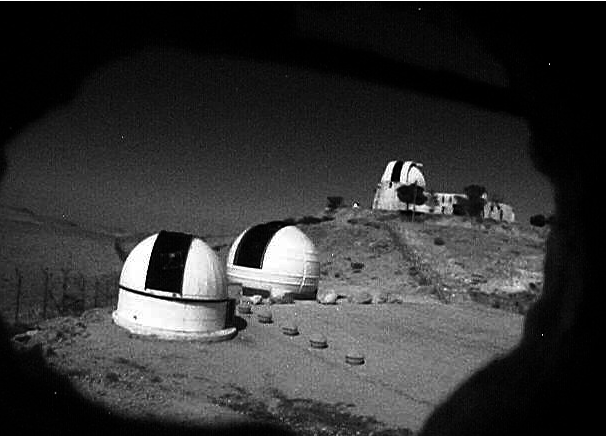}
  \caption{The dome of the 46\,cm telescope (C18) during an observing run in Spring 2016. The telescope is the first one pointing to the web-cam which is used to monitor external operations. In the background there is the dome of the Jay Baum Rich 71 cm telescope (C28), and on the hill is the dome of the 1 meter telescope of the Wise Observatory.\label{fig:telescope}}
\end{figure}

Motivated by the recent progress in photometric reverberation mapping we have started an automatized photometric monitoring of 
selected AGNs with the 46 cm telescope named C18 (Fig. \ref{fig:telescope}) of the Wise observatory in 
Israel\footnote{\url{https://physics.tau.ac.il/astrophysics/wise\_observatory}}. A sample of 27 sources with brightness V 
$<17^{m}$ was selected according to tentative continuum and emission-line time delay measurements obtained in previous works. Here 
we describe the capability, accuracy, and present the first science verification data to demonstrate the performance of our 
ongoing automatized photometric monitoring.

\begin{figure}[t!]
  \centering
  \includegraphics[angle=0,width=\columnwidth]{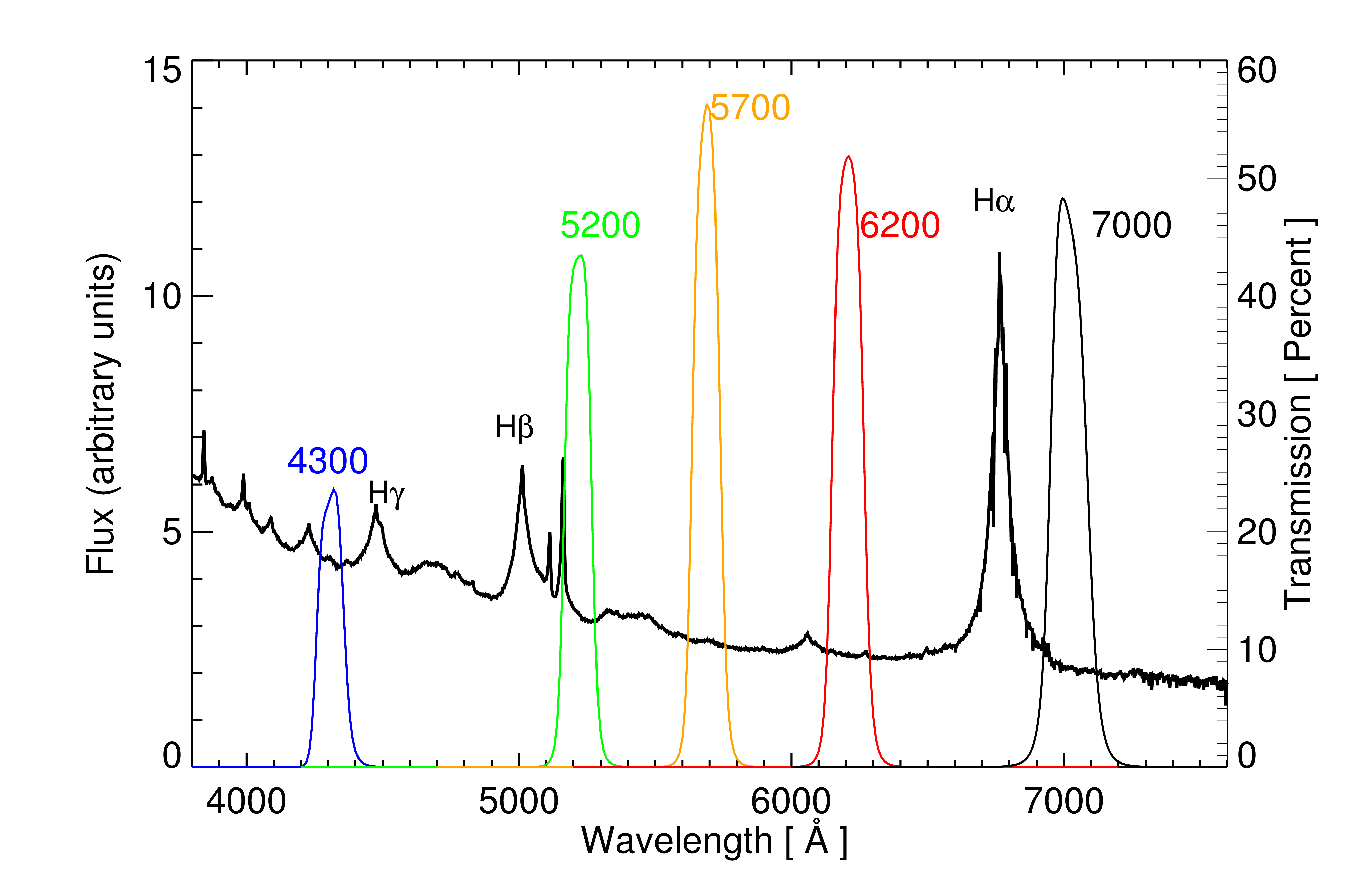}
  \caption{Composite AGN spectrum of \cite{2006ApJ...640..579G}, overlaid with the effective transmission of the narrow-band filters used for the photometric monitoring. The filters curves are convolved with the quantum efficiency of the STL-6303 CCD camera. For illustration, the AGN spectrum is shown at $z = 0.03$ where the emission line free continuum bands 4300, 5700, 6200 and 7000\,\AA\ trace mainly the AGN continuum variations, suitable for photometric reverberation mapping of the accretion disk.\label{fig:filters}}
\end{figure}

\section{Telescope and camera} \label{sec:telescope}

The telescope is equipped with a two-stage thermoelectric cooled Santa Barbara Instrument Group (SBIG) CCD STL-6303. This camera uses a Kodak Enhanced KAF-6303E/LE front-side illuminated CCD with $3072\times 2048$ $9\mu$m pixels. This provides a pixel scale of 1$\farcs$47 per pixel and a resulting field of view of approximately $75^{\prime} \times 50^{\prime}$. The addition of a water cooling system allow the CCD to operate at -$28^{\circ}$ resulting in a measured gain of 1.451 electrons ADU$^{-1}$ and a readout noise of 19.423 electrons.

\begin{deluxetable}{lcccc}
\tablecaption{Summary of filters characteristics. \label{tab:table3}}
\tablecolumns{5}
\tablenum{1}
\tablewidth{0pt}
\tablehead{
\colhead{Filter} &
\colhead{$\lambda$} &
\colhead{CWL\tablenotemark{a}} & \colhead{FWHM} & \colhead{$T(\lambda)$}
}
\startdata
NB4300 & 4260-4360 & 4311 & 100 & 21.65 \% \\
NB5200 & 5170-5260 & 5217 & 90  & 41.32 \% \\
NB5700 & 5640-5730 & 5688 & 90  & 51.83 \% \\
NB6200 & 6150-6260 & 6208 & 110 & 48.72 \% \\
NB7000 & 6950-7085 & 7018 & 140 & 44.38 \% \\
\enddata
\tablenotetext{a}{Central wavelength: $\int \lambda T(\lambda) d\lambda / T(\lambda) d\lambda$ where $\lambda$ is the wavelength 
and $T$ the filter transmission.}
\tablecomments{The effects of the detector quantum efficiency are included in the estimates. Units are in Angstroms.}
\end{deluxetable}

The camera and the telescope's optical assembly rest on an equatorial fork mount, with the right ascension (RA) and declination (DEC) aluminium disk drives being of pressure-roller types. The optical design consist of a prime-focus with a 46\,cm hyperbolic primary mirror providing a f/2.8 focus. The lightweight mirror reflects the incoming light to the focal plane through a doublet corrector lens. For more details on the technical properties of the telescope, see \cite{2008Ap&SS.314..163B}.

The STL-6303E CCD has a filter wheel with five positions and is currently equipped with five Asahi narrow-band filters with central wavelength (and FWHM) at 4300 (100), 5200 (100), 5700 (100), 6200 (120), and 7000 (120) \AA. Figure \ref{fig:filters} shows the effective transmission of the filters. The filters characteristics are summarized in Table \ref{tab:table3}.

\section{Observations} \label{sec:observations}

Observations are carried out in a robotic mode using the Astronomers Control Program (ACP\footnote{\url{http://acp.dc3.com/}}) and ACP Scheduler\footnote{\url{http://scheduler.dc3.com/}} integrated scripts. The ACP detects the weather conditions by reading the information provided by the Boltwood Cloud Sensor II weather station and the Clarity 
program\footnote{\url{http://diffractionlimited.com/}}. Once the start of observation schedule and the 
weather criterion are satisfied, the dome is automatically opened and calibration frames are taken. Usually 10 bias and dark frames 
are obtained per night. The ACP also takes at least 8 flatfields in each filter per night during dusk and dawn, adjusting the exposure time automatically from 1 to 6 seconds in order to obtain a median of about 40000 counts in all filters.

\begin{figure}[t!]
  \centering
  \includegraphics[angle=0,width=\columnwidth]{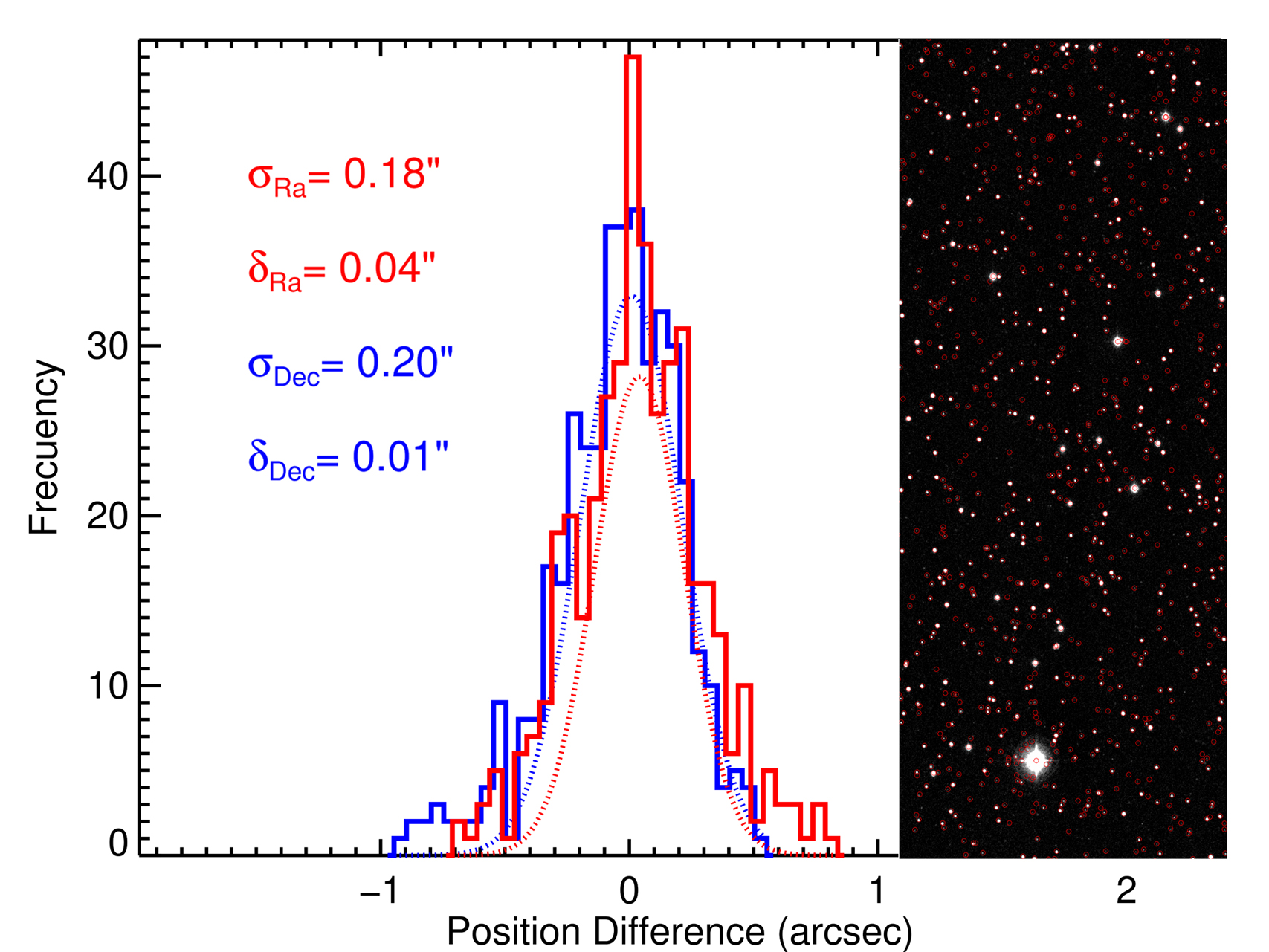}
  \caption{Astrometric precision in the field of the Seyfert 1 galaxy Mrk509. The histogram show the distribution of position differences in RA (red) and DEC (blue) for 2000 stars on the field of Mrk509 matched with the GSC2.3 catalogue. The inset shows the stars (red circles) that have been matched using a radius of 1$\farcs$ The distribution of differences in positions are well characterized by Gaussians (dotted lines) with $\sigma_{RA}=0\farcs18$ and $\sigma_{DEC}=0\farcs20$. Systematic errors ($\delta$) are typically less than $0\farcs05$ in RA and DEC.
\label{fig:astrometry}}
\end{figure}

To avoid contamination produced by hot pixels, cosmic rays and damaged sections of the CCD, we implemented a pre-defined 6 point-square dithering pattern shifted by 27.8 pixels. Astrometric solution is computed after the integration of each dithered exposure using the PinPoint\footnote{\url{http://pinpoint.dc3.com/}} astrometric engine program, which has a built-in all-sky plate solving engine including distortion mapping. Pinpoint uses the cloud solving service of Astronometry.net (\citealt{2010AJ....139.1782L}). A typical catalog-field star matching and astrometric solution process for about 700 stars takes approximately 3 seconds to complete. We have found that the astrometric precision is better than 0$\farcs$3 independent of the airmass and seeing conditions. This automatic astrometric solution has also the advantage of re-pointing the telescope in case of large pointing errors. The astrometric precision is illustrated in Figure \ref{fig:astrometry}.

\begin{figure}
  \centering
  \includegraphics[angle=0,width=\columnwidth]{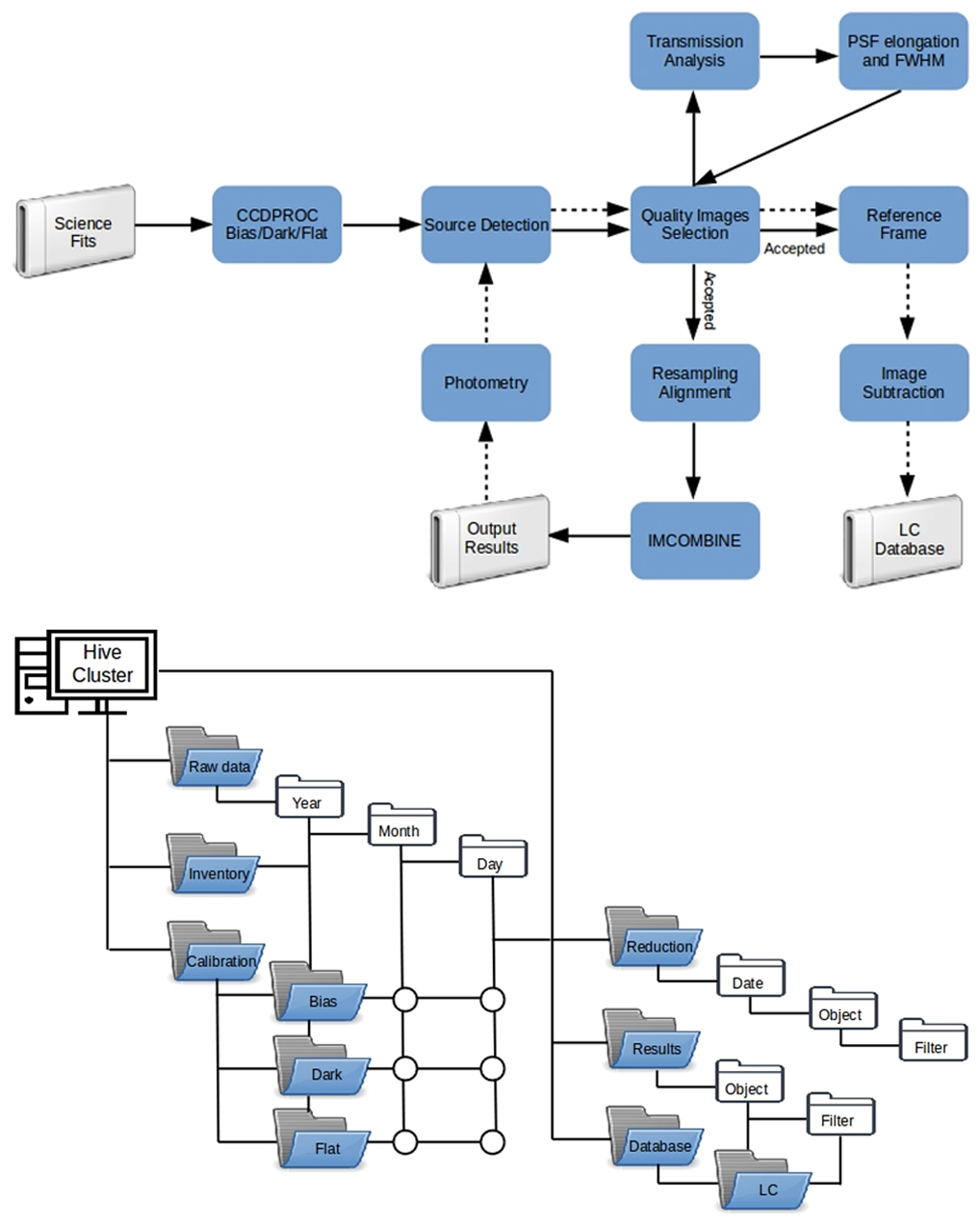}
  \caption{Flow chart illustration with the main steps of the operation of the pipeline (top). Data reduction steps are connected with a solid line, and a dotted line indicates the photometric processes (aperture photometry or image subtraction, see the text) performed on the science data product. The main structure of the data flow during the reduction process is shown in the bottom panel.\label{fig:redsteps}}
\end{figure}

During critical weather conditions caused by high humidity, strong wind or clouds, the system is able to close the dome automatically. Once conditions improve, the observations are resumed and automatically re-scheduled according to pre-defined constraints provided by the observer. This constraints involves elevation limit, moon visibility and airmass. In addition to the weather station, two independent methods for cloud detection have been implemented. Firstly, a direct analysis of the stability of the flat-fields is performed at the beginning of the observing run. Secondly, one might expect that the amount of detections obtained during the astrometric process should remain constant after consecutive exposures. Therefore large variations due to a drop in detections allow us to discriminate between cloudy and clear conditions.

\begin{figure}[t!]
  \centering
  \includegraphics[width=\columnwidth,height=8cm]{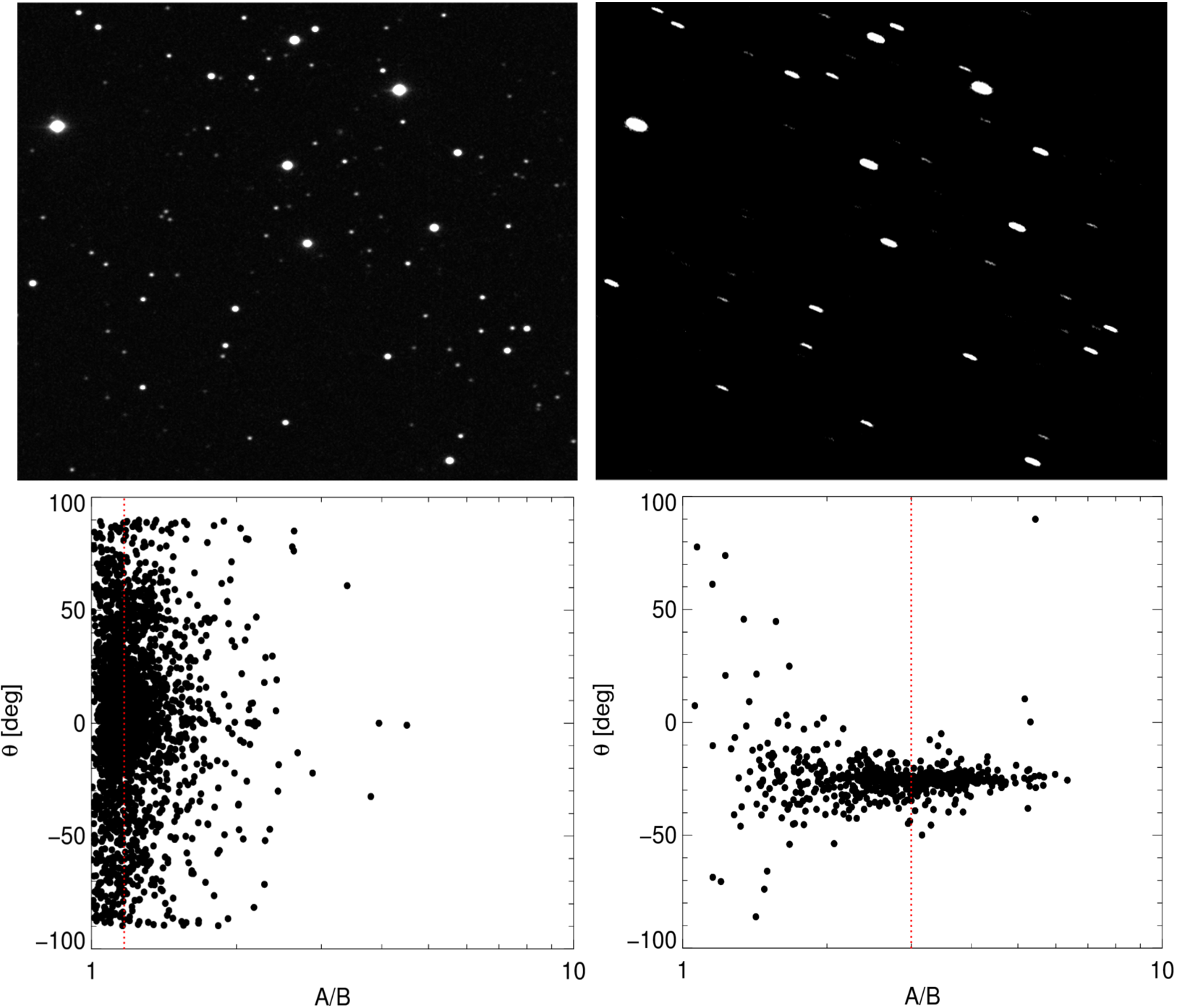}
  \caption{Two examples of the image quality selection analysis in the field of the Seyfert 1 galaxy Mrk509. The red dotted lines indicate the median of the elongation distribution. {\it Left}: An image classified as "good" with a median elongation of $1.16$ and a detection ratio of $1.08$. {\it Right}: An image classified as "bad" with a median elongation of $2.98$ and a detection ratio of $0.37$.\label{fig:quality}}
\end{figure}

\section{Data Reduction} \label{sec:data}

Data reduction is performed using our custom-made automatic pipeline, including standard IRAF\footnote{IRAF is distributed by the National Optical Astronomy Observatory, which is operated by the Association of Universities for Research in Astronomy (AURA) under cooperative agreement with the National Science Foundation.} and Astromatic\footnote{\url{http://www.astromatic.net/software/}} routines running in IDL\footnote{IDL is a trademark of Research Systems Inc., registered in the United States.} and MATLAB\footnote{\url{http://www.mathworks.com/}} environments. The pipeline organises the data science workflow, reduces the data, measures the source fluxes, and constructs light curves.
A schematic diagram of the pipeline workflow is shown in Figure \ref{fig:redsteps}. The operation of the pipeline can be described by the following basic steps:

\begin{enumerate}
\item {\bf{Data transfer and organization}}. At the end of each observing night the data are transferred to the Hive computer cluster at the University of Haifa. The images are organized using a data structure considering the year, month and day of observation. Then a catalogue for the night including all observed science and calibration images is created and stored in a database. The database is then read and the files are distributed in another data set structure that contains calibration and sciences frames. A data graph for the data set reduction process is shown in Figure \ref{fig:redsteps}.

\begin{figure*}
  \centering
  \includegraphics[width=15cm,height=7cm]{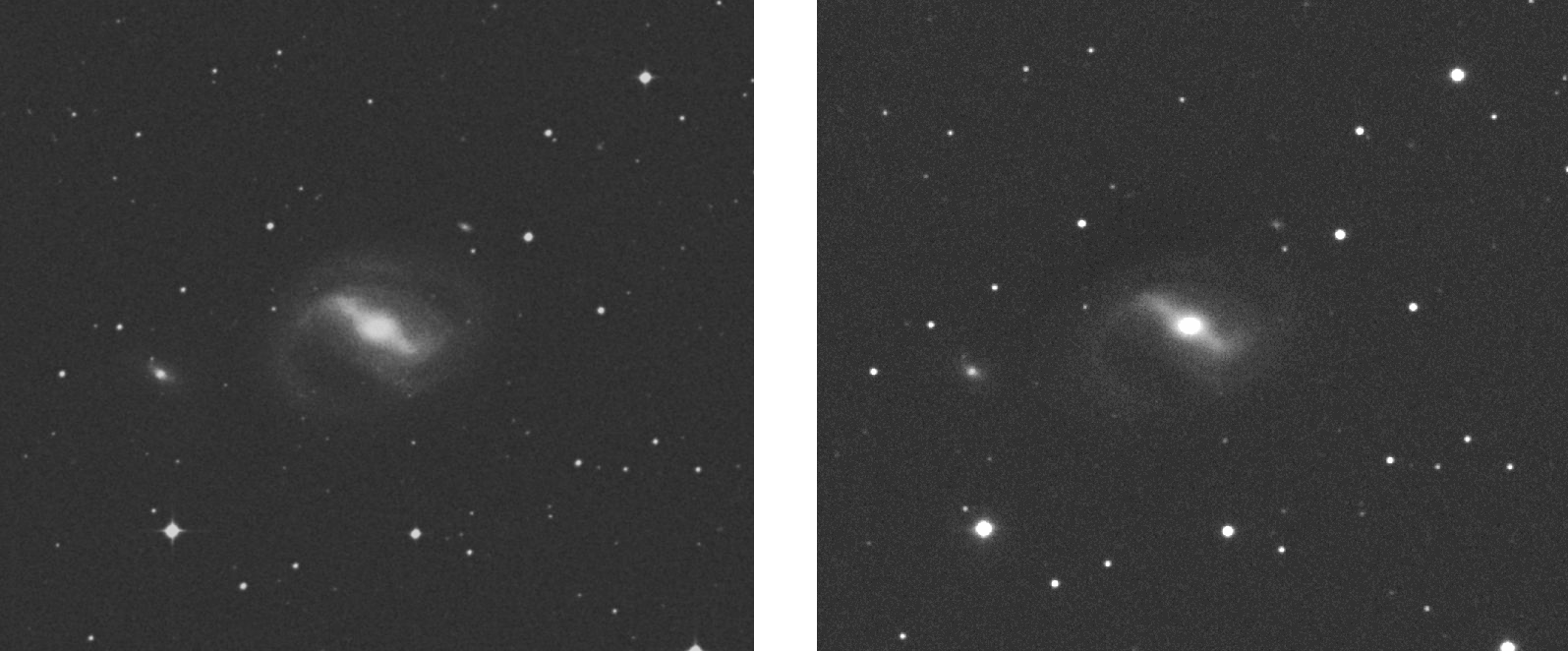}
  \caption{Images for NGC4593 from the SERC survey acquired with the 1.2 meters UK Schmidt Telescope (UKST) (left) and the 46\,cm (right). The image from the SERC survey was taken with the UKSTU red filter (IIIaF+OG590) allowing comparison with our narrow-band filter at 6200\,\AA.\label{fig:comparison}}
\end{figure*}

\item {\bf{Pre-calibration}}. Master bias, dark and flat are created daily and monthly using the IRAF tasks zero-combine, dark-combine and flat-combine. The single raw dithered frames are reduced in standard manner using the IRAF ccdproc task. 

\item {\bf{Source detection and quality image selection}}. We use SExtractor (\citealt{1996A&AS..117..393B}) to identify all the objects in the field, to estimate the fluxes and sort out the images according to their quality. Images with poor transmission and elongated stars due to tracking or auto-guider problems are classified as "bad". We use the Auto flux mode to select individual apertures and to derive the semi-major (A) and semi-minor (B) axis lengths, including the angle ($\Theta$) of the aperture relative to the horizontal-axis position of each star. Images with median elongation (A/B) larger than 1.6 and median source detection smaller than 65\% of the median distribution are automatically rejected (Fig. \ref{fig:quality}).

\item {\bf{Resampling and alignment}}. Once a good set of images is found, we apply SCAMP (\citealt{2006ASPC..351..112B}) and SWARP (\citealt{2002ASPC..281..228B}) routines to further improve the quality of the images and to accurately register the positions of the sources with respect to a reference frame. During this process, each frame is resampled from the instrumental pixel size of $1\farcs47 \times 1\farcs47$ to a new grid with a pixel size of $0\farcs75 \times 0\farcs75$ by folding the original pixels with a Lanczos-3 kernel. SWARP determines the sky background level on each iteration.

\item {\bf{Image Stacking}}. After the images have been resampled and aligned, we remove the sky background and the images are coadded using the IRAF imcombine task. We use the min/max rejection algorithm in order to remove bad pixels and cosmic rays.

\end{enumerate}
The resulting output images measure about $3600 \times 3600$ pixels and are about 50 MB in size. For comparison we show in Figure \ref{fig:comparison} the final reduced images of the field of the Seyfert 1 galaxy NGC4593 together with the same field but from the Digitized Sky Survey 2 (DSS2) obtained with the 1.2\,m Schmidt Telescope. This example demonstrates that the quality of the images obtained with our 46\,cm telescope are comparable to the quality obtained with the 1.2\,m Schmidt Telescope.

\section{Light curves} \label{sec:lc}

\begin{figure*}
  \centering
  \includegraphics[width=15cm,clip=true]{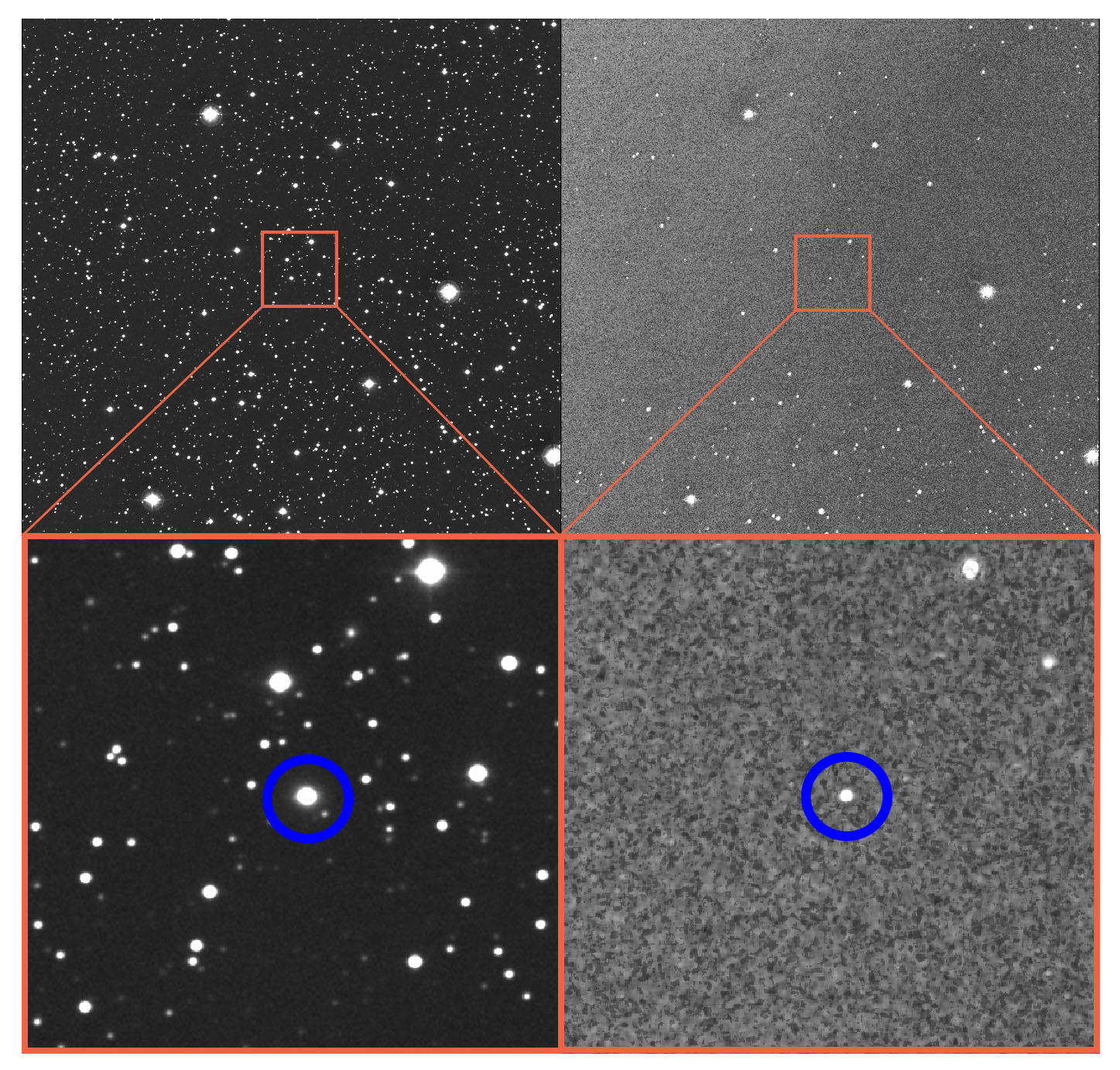}
  \caption{Image of the field containing the Seyfert 1 galaxy Mrk509. The reference image obtained with ISIS from stacking 25 best images (left) and the subtracted image for the night 2016-Oct-02 (right). The reference image contains $\sim5000$ objects from which $\sim92\%$ have been removed during the subtraction process. The bottom panels shows an area of 25$\arcmin$ square around the AGN for better visibility. The area include two faint stars close to the AGN that were successfully removed (see the text).\label{fig:sub}}
\end{figure*}

Images were analysed using image subtraction techniques based on the algorithms from \cite{1998ApJ...503..325A} subsequently generalised by \cite{2000A&AS..144..363A} implemented in the ISIS package\footnote{\url{http://www2.iap.fr/users/alard/package.html}}.
In this method a reference image is convolved with a spatially variable kernel to match the point spread function (PSF) of each individual frame. The PSF-matched images are then subtracted from each individual frame in order to separate the constant flux from the variable flux. The residual fluxes represent a difference with respect to the fluxes of the reference image. Positives fluxes represent an excess and negative fluxes a deficit from the nucleus of the AGN with respect to the flux of the reference image. 

One of the main advantages of image subtraction techniques is the removal of the extended host-galaxy contribution observed in some AGNs. The quality achieved in the subtracted images strongly depends on the quality of the kernel determination, which in turn is highly dependent on the number of stars present in the field. For some of the sources the amount of stars in the field is not enough to achieve a reliable PSF kernel determination, and therefore we extract the light curves using aperture photometry on the original images. The optimal photometric aperture is chosen to minimize the host-galaxy contribution.

The image subtraction procedures, as implemented in the ISIS package, can be summarized as follow: (1) registration of the frames to a common coordinate system, (2) construction of the reference image by stacking the best quality frames from the data set, (3) subtraction of the reference image from individual frames, (4) identification of the variable AGN, and (5) extraction of the nuclear flux from the difference images using PSF photometry.

The frames alignment procedure can be performed by ISIS itself, however we choose to register the frames using SWARP in a much earlier stage of the pipeline. The SWARP registrations make use of SExtractor for source identification and centroiding which produce better results with our images.

\begin{figure*}
  \centering
  \includegraphics[width=15cm,clip=true]{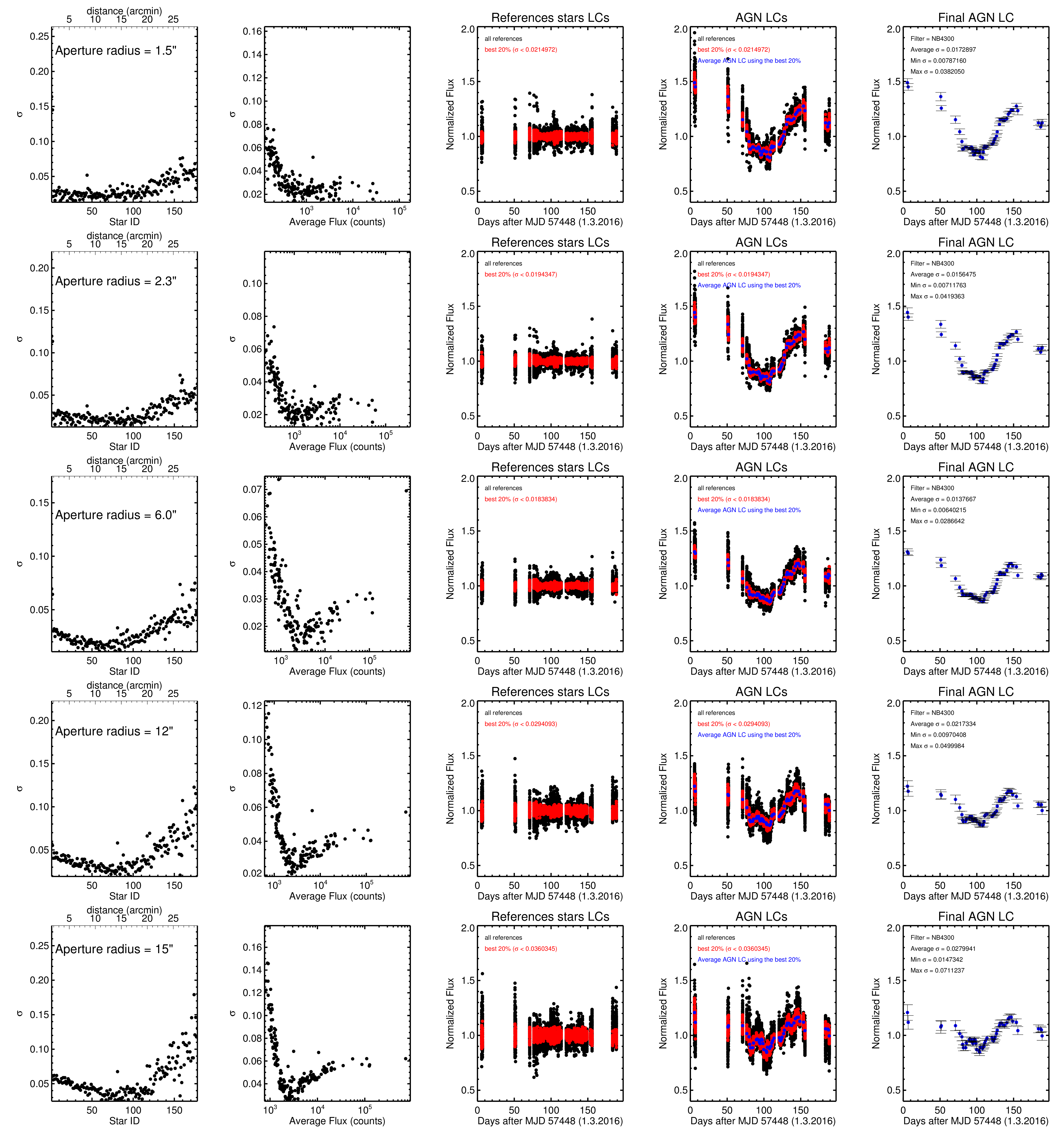}
  \caption{Photometry for the Seyfert 1 galaxy Mrk279 and the stars in the field using different apertures. From left to right: a) standard deviation ($\sigma$, in counts) of light curves for several stars in the field identified with their respective ID numbers. The top x-axis shows the distance from the center of the image. The center coincides with the position of the AGN in the image and the stars are numbered by their distance to the AGN. An increase of $\sigma$ is seen for objects located at distance larger than $\sim20$ arc-minutes. b) $\sigma$ vs stars average fluxes. c) The light curves for the reference stars (normalized flux to median = 1). Black dots are the light curves using all the stars. Red dots are the best 20\% light curves with lower sigma. d) The AGN light curve. The black dots are the AGN light curves with respect to all the reference stars, with red with respect to the 20\% with lower $\sigma$ and with blue the final average AGN light curve (using the best reference stars). e) The Final AGN light curve with the respective average, minimum and maximum $\sigma$. In this example, the best light curve which maximizes the S/N ratio and delivers the lowest scatter for the fluxes is found for an aperture of 6 arcsec.\label{fig:lcphot}}
\end{figure*}

We used ISIS default \textit{ref.csh} algorithm to build a reference image from a set of high quality frames. First, our pipeline initialises a second quality control to select the best frames for each field and filter. The analysis is based on the geometry of the PSF of each final combined image. This step is identical to the first quality control performed on the single dithered frames, as illustrated with the dotted arrows in the diagram of the pipeline work-flow shown in Figure \ref{fig:redsteps}. The selection procedure usually yields between 20 and 30 frames to combine. Provided with the best images, we use ISIS to convolve them with a spatially variable convolution kernel in order to match all the images to the same PSF. The convolved images are then stacked using a 3 sigma rejection from the median.

We next applied image subtraction to all the images using the standard ISIS \textit{subtract.csh} algorithm. The algorithm convolves the PSF of the reference frame with a spatially variable kernel to match it to the PSF of the input images before subtraction. The results are subtracted images that only contain the fluxes from variable sources while all the constant sources are removed. Figure \ref{fig:sub} illustrates the reference and difference images of the field of the Seyfert 1 galaxy Mrk509. The reference image was constructed from 25 best images in the $6200$\,\AA\, narrow-band. Note that the crowded field ($\sim200$ stars) within 5$\arcmin$ around the AGN has been removed during the subtraction process, and only two stars remain which are classified as variables. Particularly interesting are also the two stars with magnitudes $B\sim19.8$ and $B\sim20.1$ separated only $15$ arcsec from the AGN. The results illustrate the high accuracy in the detection and subtraction even for very faint sources.

At this stage, the ISIS algorithm \textit{detect.csh} can be used to detect only the variable sources on each subtracted frame. However, we prefer to use SExtractor on the reference frame to produce a catalogue that include the variable AGN and all the references stars in the field. The residual fluxes of all the constant stars are used as an estimate of the uncertainty in the subtraction process. Then, we feed the ISIS PSF photometric algorithm \textit{phot.csh} with the catalogue obtained with SExtractor and create light curves for the AGN and the reference stars.

\begin{figure}
  \centering
  \includegraphics[width=\columnwidth]{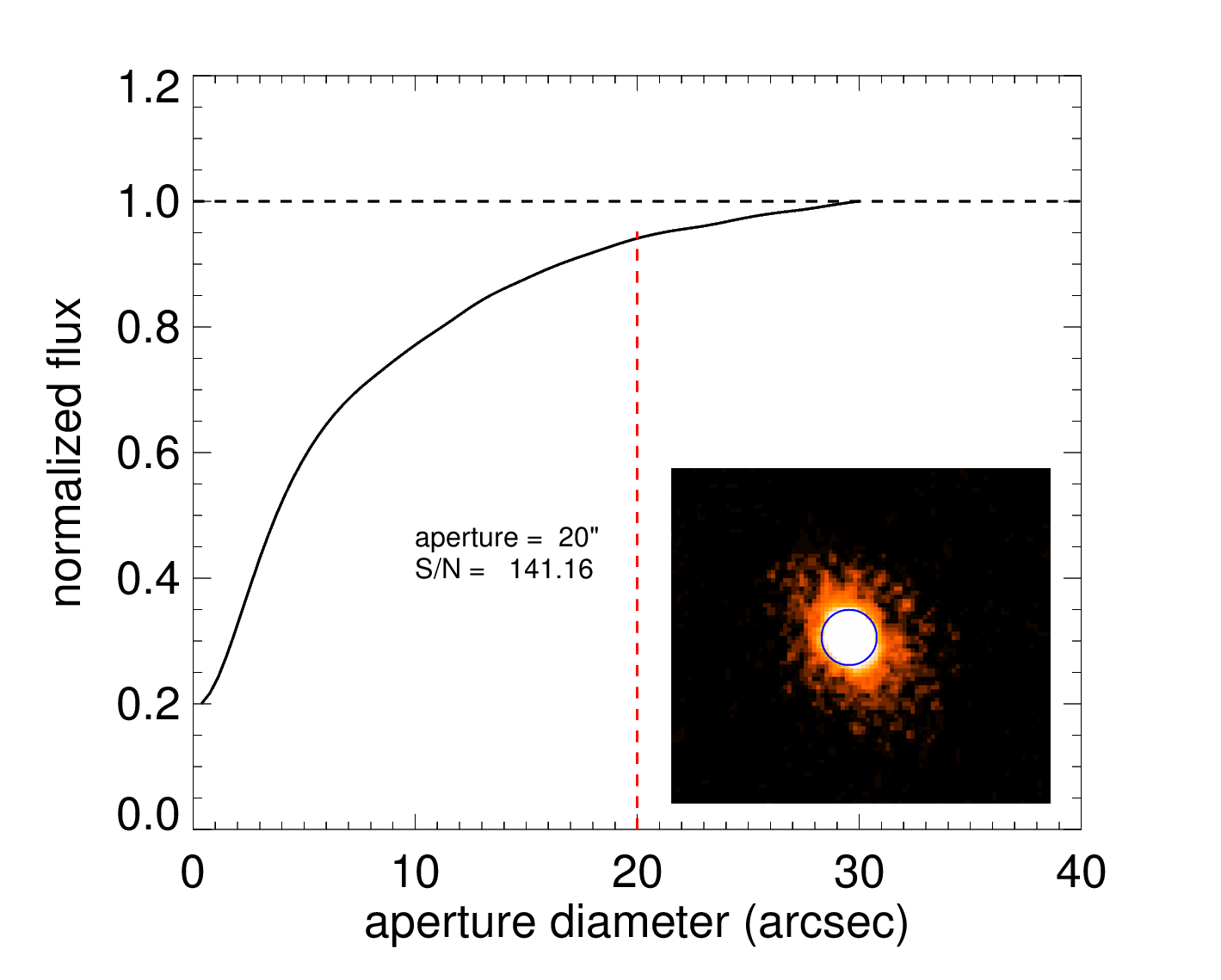}
   \caption{Illustration of a photometric curve-of-growth obtained for the Seyfert 1 galaxy NGC7469. The inset shows an image of the AGN and the extended (45 arcsec) host-galaxy. We found that an aperture of 20 arcsec around the nucleus (blue circle) maximizes the signal-to-noise ratio (S/N = 141).\label{fig:cofgro}}
\end{figure}

\begin{figure}[t!]
  \centering
  \includegraphics[width=\columnwidth]{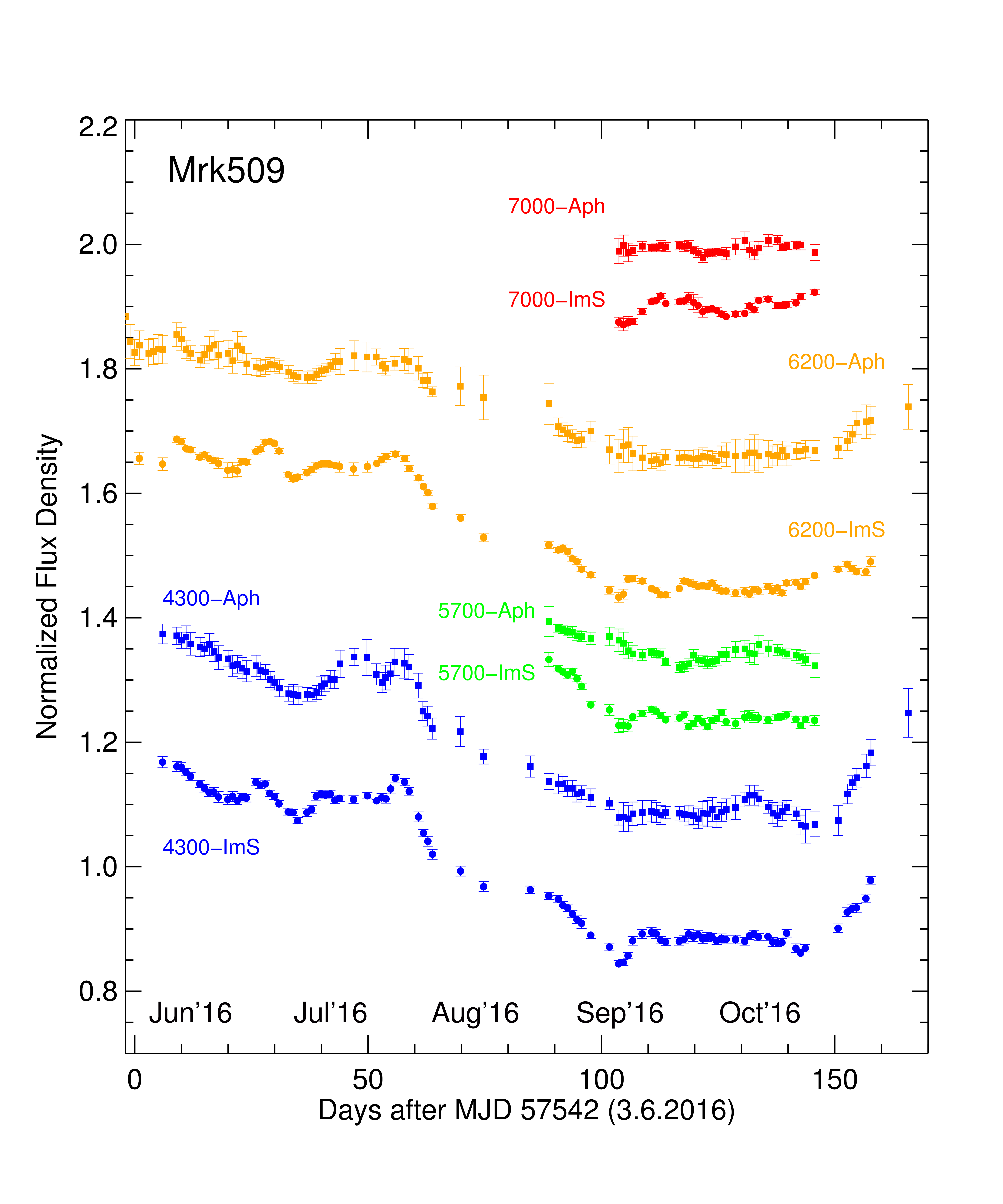}
   \caption{Image subtraction (ImS) and aperture photometry (Aph) light curves for Mrk509 obtained between Jun. 2016 and Oct. 2016. The light curves are normalised to their mean and vertically shifted for clarity.\label{fig:aphims}}
\end{figure}

For AGN fields that do not have enough bright stars to reliable determine a PSF kernel solution, aperture photometry is applied directly on the reduced images. First, in order to quantify the aperture that maximizes the signal-to-noise ratio (S/N) and delivers the lowest scatter for the fluxes, photometry is extracted for all the stars in the field using different apertures\footnote{Apertures between 0.75 and 23 arcsec are typically used. An extended version of this figure can be found in \url{http://www.pozonunez.de/figure8_p1.html}} (Fig.\ref{fig:lcphot}). Second, a photometric curve-of-growth is constructed for the AGN in order to compare and trace the effects of the host-galaxy and AGN contribution (Fig. \ref{fig:cofgro}). Third, the sources are categorized according to the distance from the AGN, brightness and their light curves standard deviations ($\sigma$). Using very small apertures (1.5 and 2.3 arcsec) yields larger $\sigma$. This effect is expected because we are only considering a small portion of the total flux which remains heavily dependent on the quality of the PSF. Similarly, $\sigma$ increases for very large apertures (12 and 15 arcsec) where the results are more sensitive to the sky background contamination. We find that usually $\sigma$ of the reference stars depends on the brightness. We therefore select the reference stars with similar brightness as the AGNs and located within 20$\arcmin$ around the AGNs in order to avoid possible systematics across the field introduced during the reduction process. We chose as reference stars the best 20\% of the stars with the lowest $\sigma$. This selection procedure usually yields about 15 to 30 stars. We create relative light curves using the previous selected reference stars. The resulting average light curve obtained using the selected reference stars correspond to the final AGN light curve.

In Figure \ref{fig:aphims}, we compare the light curves obtained with image subtraction and aperture photometry for the Seyfert 1 galaxy Mrk509. While the overall shape of the light curves for different bands obtained with both photometric methods is similar, the photometric precision obtained with image subtraction (0.5-1\%) is superior to the obtained with aperture photometry (1.2-2\%). As we mentioned above, the image subtraction performance strongly depends on the quality of the PSF kernel and the amount of stars on the field. The field of Mrk509 contains $\sim5000$ stars allowing a proper PSF kernel determination and high-quality subtraction. One possible explanation for the different photometric precision is that the optimized photometric aperture might enclose a larger contribution from the extended host-galaxy ($\sim15"$) leading to higher uncertainties. On the other hand, the constant host-galaxy contribution has been successfully removed during the subtraction process and only the variable nuclei stand out in the subtracted images. 

\begin{figure}
  \centering
  \includegraphics[width=\columnwidth]{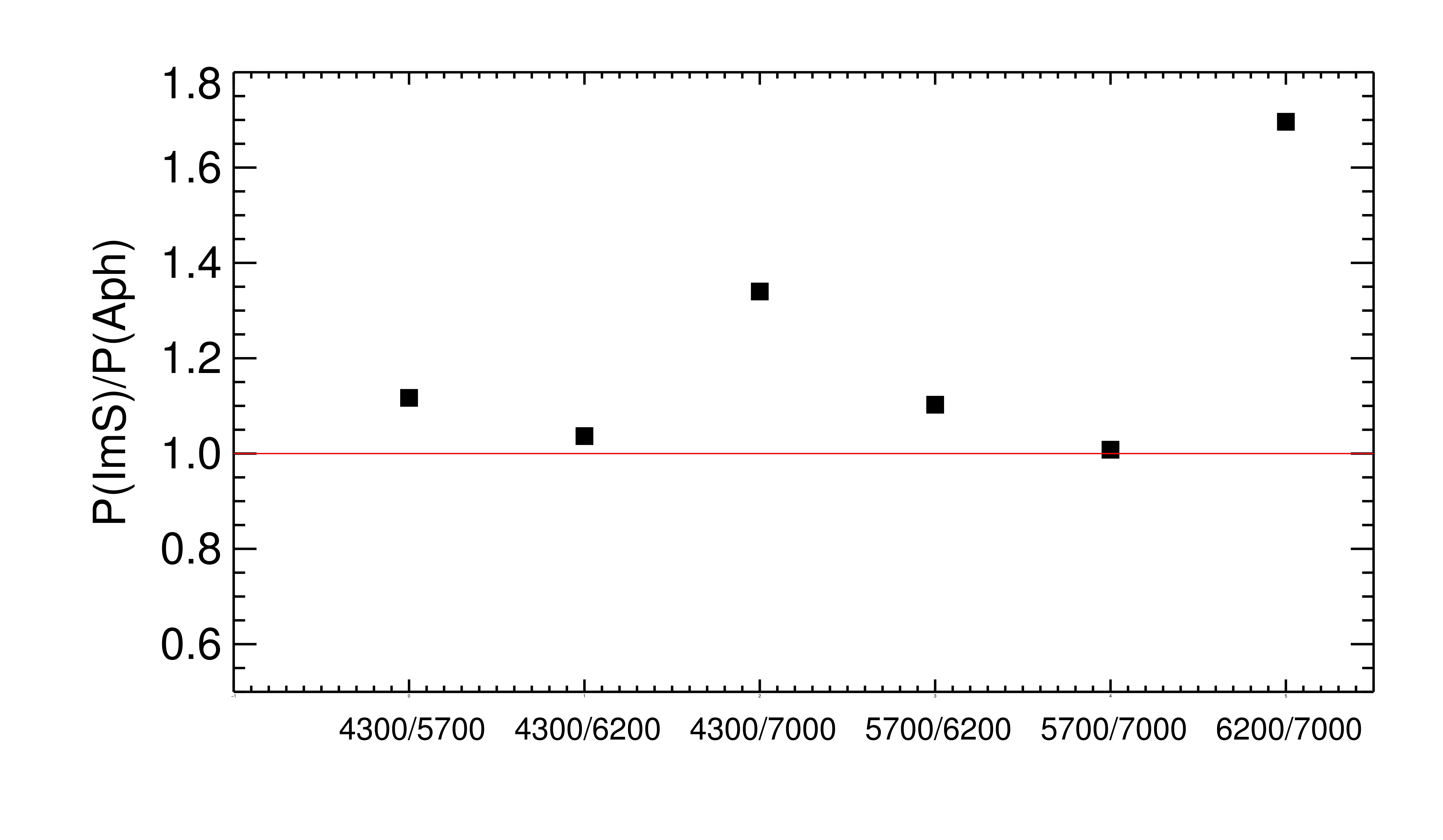}
   \caption{Pearson correlation coefficient between the light curves in different bands obtained with image subtraction and aperture photometry. $P > 1$ implies a higher correlation of the light curves obtained with image subtraction.\label{fig:pearson}}
\end{figure}

In order to further quantify the particular performance of both methods, we compute the Pearson correlation coefficient between the light curves. Without considering the effects of time delays one might expect that the continuum emission between the bands should be highly correlated. We define the ratio $P = P_{ImS}/P_{Aph}$ where $P_{ImS}$ and $P_{Aph}$ are the Pearson correlation coefficients for image subtraction and aperture photometry, respectively. There is a moderate correlation between the light curves obtained with both methods. However, the Pearson correlation coefficient obtained for image subtraction is always higher than for aperture photometry in all bands ($P_{ImS}/P_{Aph} > 1$), as can be seen in Fig. \ref{fig:pearson}. We conclude that image subtraction outperform aperture photometry for the particular case of Mrk509. A more detailed exploration of the performance of both methods is beyond the scope of the present paper.

\begin{figure}[t!]
  \centering
  \includegraphics[width=\columnwidth]{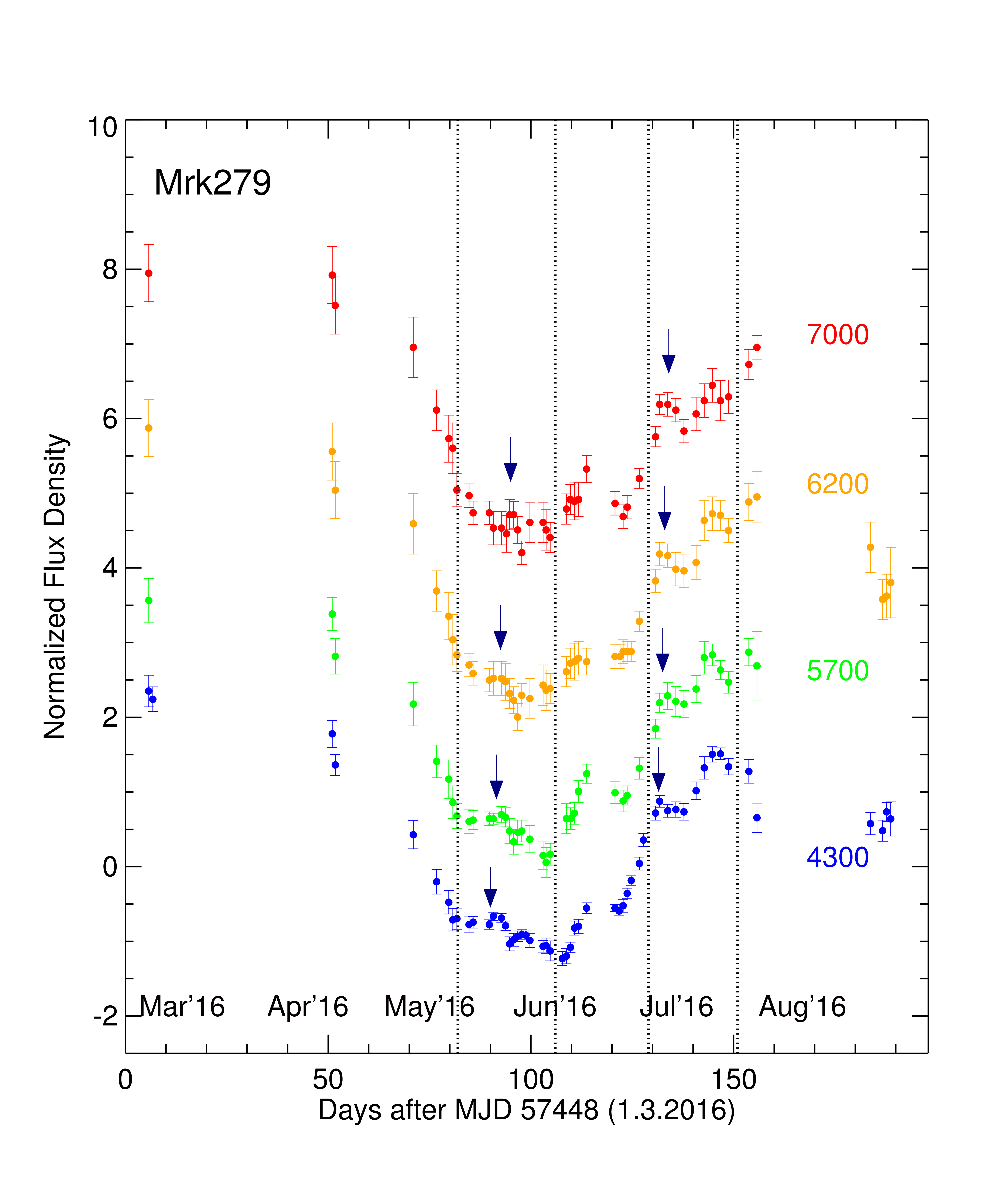}
   \caption{Observed light curves of the Seyfert 1 galaxy Mrk279 for the period between March 2016 and August 2016. The light curves are normalized to zero mean, unit standard deviation and vertically shifted for clarity. The dotted lines indicates two different windows of short variability events. The arrows mark the variation features for which a possible time delay can be detected by eye (see the text). \label{fig:lcmrk}}
\end{figure}

In the followings we illustrate the results for the Seyfert 1 galaxy Mrk279. Through Mrk279 redshift of $z = 0.0306$, the emission line free continuum bands 4300, 5700, 6200 and 7000\,\AA\, trace mainly the AGN continuum variations (Fig. \ref{fig:filters}). Consequently, it is an ideal candidate for photometric reverberation mapping of the accretion disk. The light curves of Mrk279 are shown in Figure \ref{fig:lcmrk}. The bands shows a gradual flux decrease by about $40\%$ from the beginning of March until a minimum is reached at the beginning of June. Afterwards, the flux undergoes a steep increase by about the same amplitude ($40\%$) until a maximum is reached at the end of July. The five days of observations between the end of July and the beginning of September show a more regular variability. The poor time sampling achieved between the beginning of March and end of April, and including the end of the campaign is due to the weather conditions preventing us from opening the dome. However, we preferred to include them in the light curves in order to illustrate the state of the AGN before and after the strong period of variability events. The low altitude of the source at the end of September only allowed us to have observations in the 4300 and 6200\,\AA\, bands. While the long-term variations seems to be well correlated between different bands, the high quality of the light curves allows us to detect significant short-time scale variations, which at first glance appear to be uncorrelated. For instance, the variation features observed during the variability windows MJD57448+82-MJD57448+108, and MJD57448+129-MJD57448+151, allow us to detect by eye a time delay between the continuum bands of about 2 days. Both 4300-band steep declines starting at MJD57448+90 and MJD57448+131 are roughly consistent with the 5700-band shifted by about 1.5 days. The same holds for the 6200 and 7000-bands during both variability windows (marked by the dotted lines and arrows in Fig. \ref{fig:lcmrk}). A more detailed study of the light curves of Mrk279, including other AGNs from the campaign will be presented in a forthcoming publication.

\begin{figure}[t!]
  \centering
  \includegraphics[angle=0,width=\columnwidth]{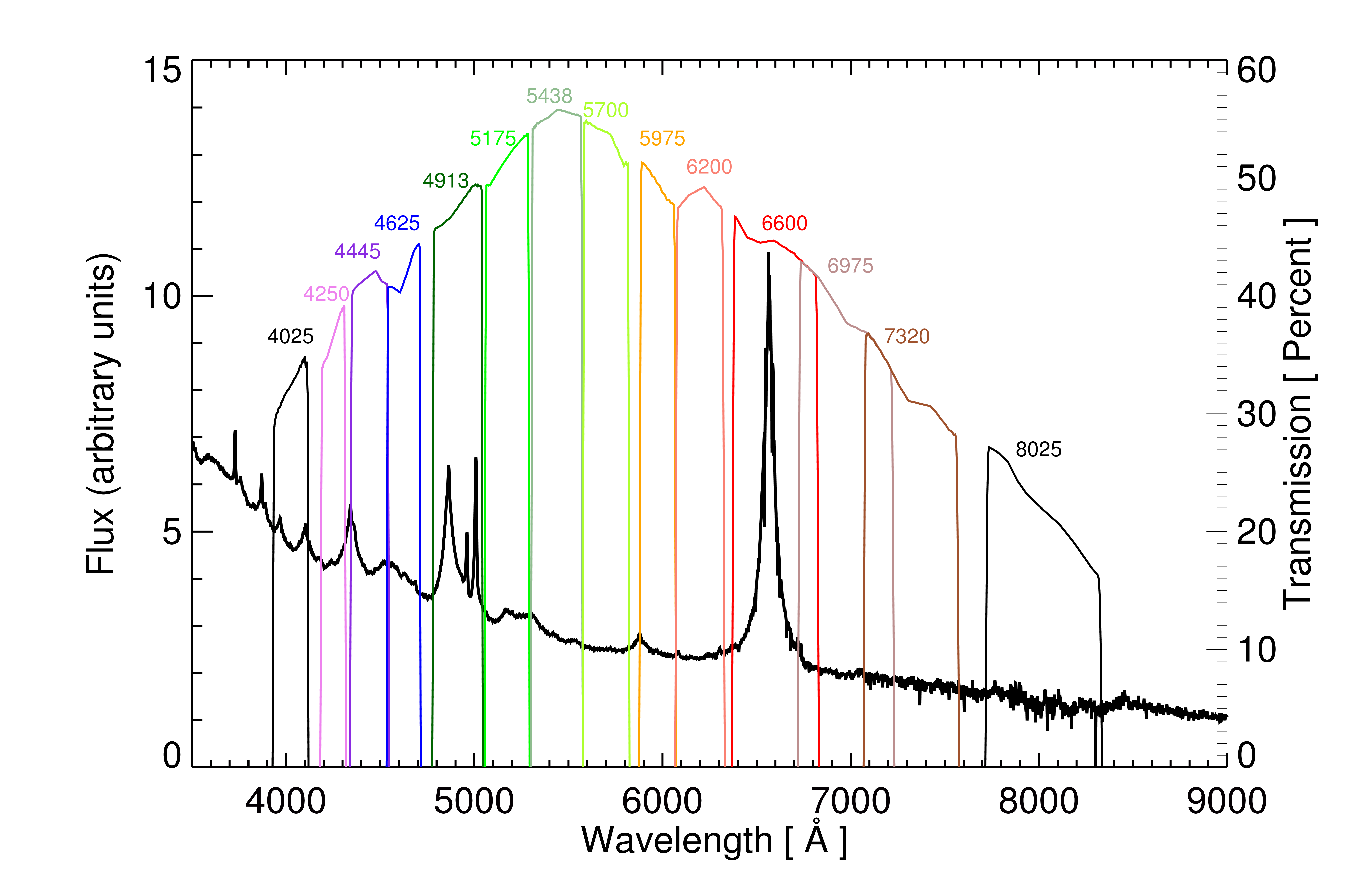}
  \caption{Same as Fig. \ref{fig:filters}, but for the new commissioned 14 filters overlaid with the $z = 0$ composite AGN spectrum of \cite{2006ApJ...640..579G}. The filters curves are convolved with the quantum efficiency of the QSI KAF-8300 CCD camera. The numbers on top of the filters transmissions correspond to the central wavelengths.\label{fig:nextfilt}}
\end{figure}

\section{Summary and outlook} \label{sec:summary}

We have presented the first results of a new photometric reverberation mapping monitoring of AGNs, using the 46\,cm telescope of the Wise observatory in Israel. We have described the automatic observations and data reduction steps necessary to achieve efficient unassisted observations and high-quality light curves. The use of image subtraction and aperture photometry techniques allows us to measure the nuclear flux of the AGNs with high accuracy ($< 1\%$). The high precision obtained will provide also the possibility to study other variability phenomena such as variable stars in the field of the AGNs, for instance the analysis of faint stars which are subjects to more contamination from the sky background and bright stars which are saturated with larger telescopes. 

Finally, we are commissioning a second CCD camera (QSI KAF-8300) with 14 new filters. The new filters will provide a better wavelength coverage of both line-free continuum regions and BLR H$\alpha$/H$\beta$ emission lines (Fig. \ref{fig:nextfilt}), hence decreasing the biases in the time-lag measurements of the AD and the BLR for several AGNs observed in the future.

\acknowledgments

This research has been partly supported by grants 950/15 from the Israeli Science Foundation (ISF) and 3555/14-1 from the Deutsche Forschungsgemeinschaft (DFG). Hive computer cluster at the university of Haifa is partly supported by ISF grant 2155/15. 
This work is based on observations collected at the Wise Observatory with the C18 telescope. The C18 telescope and most of its equipment were acquired with a grant from the Israel Space Agency (ISA) to operate a Near-Earth Asteroid Knowledge Center at Tel Aviv University. We thank Prof. Dan Maoz, the director of the Wise Observatory, for consistent support of the project. The observations at Wise observatory benefited from the continuous support of Tel-Aviv university and from the care of the site manager Sami Ben-Gigi. 
We thank our referee Robert Antonucci for helpful comments and careful review of the manuscript. F.P.N thanks Christoph Bruckman and Angie Barr Dominguez for fruitful discussions and important feedback. This research has made use of the NASA/IPAC Extragalactic Database (NED) which is operated by the Jet Propulsion Laboratory, California Institute of Technology, under contract with the National Aeronautics and Space Administration. This research has made use of the SIMBAD database, operated at CDS, Strasbourg, France.

\bibliographystyle{aasjournal} 
\bibliography{agnwise}



\end{document}